\documentclass[12pt]{article}

\usepackage{amscd, amsmath}
\usepackage{amsthm, amssymb}
\usepackage{array}

\newcommand{\CC}{\mathbb{C}}

\newcommand{\dd}{{\rm d}}
\newcommand{\is}{I}
\newcommand{\ais}{\mathcal I}
\newcommand{\mais}{\varphi}

\newtheorem{thm}{Theorem}[section]
\newtheorem{prop}[thm]{Proposition}
\newtheorem{cor}[thm]{Corollary}
\newtheorem{lem}[thm]{Lemma}
\newtheorem{defin}[thm]{Definition}

\theoremstyle{remark}

\theoremstyle{remark}

\theoremstyle{remark}

\newenvironment*{prooff}{\noindent {\bf Proof.}}{\hfill $\qed$ \vspace{.3cm}}
\newenvironment*{prooff1.1}{\noindent {\bf Proof of theorem \ref{thm1.1}.}}{\hfill $\qed$ \vspace{.3cm}}

\begin{document}

\begin{titlepage}
\title{
\vskip -70pt
\begin{flushright}
{\normalsize \ ITFA-2008-41}\\
\end{flushright}
\vskip 45pt
{\bf Non-abelian vortices on compact Riemann surfaces}
}
\vspace{3cm}

\author{{J. M. Baptista} \thanks{ e-mail address:
    j.m.baptista@uva.nl}  \\
{\normalsize {\sl Institute for Theoretical Physics} \thanks{ address: Valckenierstraat 65, 1018 XE Amsterdam, The Netherlands}
} \\
{\normalsize {\sl University of Amsterdam}} 
}

\date{October 2008}

\maketitle

\thispagestyle{empty}
\vspace{1.5cm}
\vskip 20pt
{\centerline{{\large \bf{Abstract}}}}
\vspace{.2cm}
We consider the vortex equations for a $U(n)$ gauge field $A$ coupled to a Higgs field $\phi$ with values on the $n\times n$ matrices. It is known that when these equations are defined on a compact Riemann surface $\Sigma$, their moduli space of solutions is closely related to a moduli space of $\tau$-stable holomorphic $n$-pairs on that surface. Using this fact and a local factorization result for the matrix $\phi$, we show that the vortex solutions are entirely characterized by the location in $\Sigma$ of the zeros of $\det \phi$ and by the choice of a vortex internal structure at each of these zeros. We describe explicitly the vortex internal spaces and show that they are compact and connected spaces.

\vspace{.45cm}

\end{titlepage}

\section{Introduction}

\subsection{The context}

The simplest and earliest-known type of vortex equations appeared in the classical abelian Higgs model. It involves just one $U(1)$ gauge field and one complex scalar Higgs field. The solutions to these equations are generally  known as Nielsen-Olesen vortices \cite{A-N-O}, and the moduli space of such solutions was first described by Taubes for vortices living in the complex plane \cite{Tau}, and by Bradlow for vortices living in a compact K\"ahler manifold \cite{Brad}. After that several far-reaching generalizations of the vortex equations have been studied. In one of these the gauge group can be any compact connected Lie group and the Higgs field can have values in any K\"ahler manifold equipped with a hamiltonian action of the group. Among these generalizations, perhaps the simplest ones to deal with and explicitly describe solutions, are the ones that have a torus as the gauge group and a Higgs field with values on a toric manifold \cite{Bap}. These are abelian generalizations, of course. The non-abelian generalizations, in their turn, are more difficult to analyse, but at the same time seem to possess a richer structure and present the greatest number of novel features, as for instance the presence of vortex internal spaces. A great deal of effort has therefore been dedicated to studying various types of non-abelian vortex equations, and we suggest for example \cite{B-D-GP-W, Tong} for succinct reviews --- one from a mathematical and one from a physical perspective --- and for references to the many original articles.

In this paper our intent is to focus on what is probably the simplest non-abelian model for vortices: that with one $U(n)$ gauge field and one linear Higgs field having values on the space of complex $n \times n$ square matrices. We will study these vortex equations when they are defined on compact Riemann surfaces of large volume and will give an explicit and rigorous description of their moduli space of solutions. Since an apparent feature of the literature is that the existent rigorous accounts of non-abelian vortex moduli spaces seem to be rather hard to be made explicit, and vice-versa, we thought that this study could be of some interest.

The data that we need to start with are a compact Riemann surface $\Sigma$ and a complex vector bundle $E\rightarrow \Sigma$ of rank $n$ over that surface. Recall that, as a $C^\infty$ vector bundle, $E$ is completely characterized by its degree $d$, which we assume to be positive. Fixing an hermitian metric $h$ on $E$, the variables of our field theory are then the unitary connections $A$ on the bundle and the sections $\phi$ of the direct sum $\oplus^n E \rightarrow \Sigma$ of $n$ copies of $E$. Observe that locally $\phi$ can be regarded as a function on $\Sigma$ with values on the complex $n\times n$ matrices. The fact that these matrices are square introduces significant simplifications to the problem, for one can take inverses and determinants at will, just as in the $n=1$ abelian case; it is nevertheless possible to go a long way in studying the same non-abelian model with a different number of copies of $E$ \cite{B-D-W, Eto-2}.

The energy functional of the model is the natural Yang-Mills-Higgs functional
\begin{equation}
E(A ,\phi) \ = \ \int_\Sigma \: \frac{1}{2 e^2}\, |F_A|^2 \: +\: |\dd^A \phi|^2 \: +\:  \frac{e^2}{2}\, |\phi\, \phi^\dagger -  \tau \, 1 |^2  \ ,
\label{1.1}
\end{equation} 
where $F_A$ is the curvature of the connection, $\dd^A \phi$ is the covariant derivative and $e$ and $\tau$ are positive real parameters of the theory. (Here the hermitian conjugate $\phi^\dagger$ is of course defined with respect to the hermitian metric $h$ on $E$, so that in a unitary trivialization of $E$ it is represented by the hermitian conjugate matrix of $\phi$.) As is well known a Bogomolny-type argument then shows that this energy is minimized by the fields $(A,\phi)$ that solve the vortex equations
\begin{align}
&\bar{\partial}^A \phi \ = \ 0    \label{1.2} \\
&\ast F_A - i e^2\, (\phi\, \phi^\dagger - \tau \, 1) \ = \ 0 \ ,\nonumber
\end{align}
where $\bar{\partial}^A \phi$ is the anti-holomorphic part of the covariant derivative and $\ast$ is the Hodge operator on $\Sigma$. If these equations have any solutions at all, then the energy functional at this minimum will have the value $E(A, \phi) = 2 \pi \tau d$.


The vortex equations as written in (\ref{1.2}) were first studied in \cite{B-D-W}, where they were related to the problem of finding $\tau$-stable holomorphic $n$-pairs on $\Sigma$. That paper was part of a much wider effort to analyse various types of non-abelian vortex equations and, through the use of Hitchin-Kobayashi correspondences, relate them to various types of stability conditions for vector bundles equipped with sections \cite{B-D-GP-W}.

A second wave of interest, this time in the physics literature, came after the articles \cite{H-T, A-B-E-K-Y} arrived independently at the same non-abelian equations. The first one gave a brane-theoretical description of the vortex moduli spaces; the second was concerned with their applications to confinement in QCD. There was then a sequence of articles providing alternative and more direct constructions of these moduli spaces, other related non-abelian moduli spaces and studying their physical implications (closer to the perspective of this paper, see for example \cite{Eto-1, A-S-Y, Eto-4}, but otherwise also the many others referred to in \cite{Tong}). These constructions were carried out mostly for vortices in the complex plane $\Sigma = \CC$, and while this choice introduces several topological simplifications, it also prevents the direct use of the Hitchin-Kobayashi correspondences of \cite{B-D-W, B-D-GP-W}, for these have so far been proved only for compact $\Sigma$. This means that for $\Sigma =\CC$ the moduli space constructions in the physical literature are not yet completely rigorous, though they certainly are very useful and almost surely true. In the present paper we introduce a novel way to characterize non-abelian vortex solutions --- in terms of the locations and internal structures of the zeros of $\det \phi$ --- which is applicable to both compact and non-compact $\Sigma$. In the compact case we can then make use of the correspondences of \cite{B-D-W, B-D-GP-W} to confidently describe the vortex moduli spaces.

\subsection{The main result}

We now describe the main result of the paper. As a first point, observe that associated to $E$ there is the natural $C^\infty$ complex line-bundle $\det E \rightarrow \Sigma$. This determinant bundle also has degree $d$, its transition functions are the determinants of the transition functions of $E$, and a complex structure on $E$ induces a complex structure on $\det E$. Now, if $(A, \phi)$ is a solution of the vortex equations, then it is well known that the connection $A$ determines a complex structure on $E$ such that the section $\phi : \Sigma \rightarrow \oplus^n E $ becomes holomorphic \cite{B-D-GP-W}. This follows from the first vortex equation. This $\phi$ then determines a holomorphic section $\det \phi$ of the determinant bundle $\det E$ with the induced complex structure. But being holomorphic, the section $\det \phi$ either vanishes everywhere or has exactly $d$ zeros, counting multiplicities. In this paper we concentrate on the latter case, i.e. on vortex solutions such that $\det \phi$ does not vanish identically.

Now suppose that $z_j \in \Sigma$ is one of these isolated points where $\det \phi$ vanishes. Making use of local trivializations of $\oplus^n E$ that are holomorphic with respect to the complex structure induced by $A$, the section $\phi$ can be regarded as a holomorphic function around $z_j$ with values on the $n\times n$ square matrices. We want to characterize the behaviour of $\phi$ around the point $z_j$ where the determinant vanishes, and for this we introduce the following two definitions. 
\begin{defin}
A vortex internal structure $\is_n$ is a set of data consisting of an integer $k_0 \geq 0$ and a sequence $(V_1, \ldots , V_l)$ of non-zero proper subspaces of $\CC^n$ such that $V_{j+1} \cap V_{j}^{\perp} = \{ 0 \}$ for all indices $j=1, \ldots , l-1$. The order of the internal structure $\is_n$ is the non-negative integer $n\, k_0 + \sum_l  {\rm dim}_{\CC} V_l$.
\label{dfn2.2}
\end{defin}
\begin{defin}
Given a subspace $V$ of $\CC^n$ consider the orthogonal decomposition $\CC^n = V \oplus V^{\perp}$. Calling $\Pi_V$ and $\Pi_V^{\perp}$ the associated projections, for any complex scalar $z$ one defines the elementary linear transformation 
\[
T_V (z) \ := \ z \Pi_V + \Pi_V^{\perp} \ : \ \CC^n \longrightarrow \CC^n \ .
\] 
It is clear that the determinant of $T_V (z)$ is $z^{{\rm dim}\; V}$ and that, for $z\neq 0$, the inverse $T_V (z)^{-1}$ is $z^{-1} \Pi_V + \Pi_V^{\perp}$.
\label{dfn1.1}
\end{defin}
The use of these definitions, and a key point in our results, is that the matrix function $\phi (z)$ can then be uniquely factorized around $z_j$ as
\begin{equation}
\phi (z) \ = \ A(z)\: (z-z_j)^{k_0} \; T_{V_l}(z-z_j) \cdots T_{V_1} (z-z_j)
\label{1.3}
\end{equation}
for some internal structure $\is_n  = (k_0 , V_1 , \ldots , V_l)$ with order equal to the multiplicity of the vanishing of $\det \phi$ at the point $z_j$. Here $A(z)$ is some holomorphic matrix function that is invertible around $z_j$. Since the structure $\is_n$ is independent of the chosen trivialization of $E$, we thus have a canonical way to associate to each zero of $\det \phi$ a correspondent algebraic internal structure. It then turns out that these internal structures completely determine the vortex solution up to gauge transformations. More precisely we have the folllowing result.
\begin{thm}
Let $E \rightarrow \Sigma$ be a complex vector bundle of rank $n$ and degree $d$ over a compact Riemann surface, and assume that $({\rm Vol}\; \Sigma) > 2\pi d / (e^2\, \tau)$. Now pick any finite set $\{ (z_1 , \is_n^1), \ldots , (z_r , \is_n^r) \}$ of distinct points on the surface and associated internal structures such that $\sum_{l=1}^{r} {\rm order}( \is_n^l) = d$. Then there is a solution $(A,\phi)$ of the non-abelian vortex equations (\ref{1.2}), unique up to gauge equivalence, such that $\det \phi$ has zeros exactly at the points $z_j$ and $\phi$ factorizes around each $z_j$ with internal structure $\is_n^j$. Furthermore, all solutions of (\ref{1.2}) with $\det \phi$ not identically zero are obtained in this way.  
\label{thm1.1}
\end{thm}
The condition of large volume of $\Sigma$ is required in the $\tau$-stability results of \cite{B-D-W, B-D-GP-W}, which are essential in our proof of the theorem above. At the same time, observe that a simple integration over $\Sigma$ of the second vortex equation shows that no solutions exist if $({\rm Vol}\; \Sigma)\: < \: 2\pi d / (n\, e^2\, \tau)$. Thus the general picture that arises is that for small volumes of $\Sigma$ there are no vortex solutions; then for ${\rm Vol}\; \Sigma$ in the interval between $ 2\pi d / (n\, e^2\, \tau)$ and $ 2\pi d / (e^2\, \tau)$ there is a less well known, and possibly complicated, moduli space of solutions; and finally for large volumes of $\Sigma$ (or big $e^2$, or big $\tau$) the moduli space can be neatly described as above.  

The layout of the paper is the following. In section 2 we study holomorphic matrix functions on the plane, proving the local factorization (\ref{1.3}) and a generalization thereof. In section 3 we extend this to holomorphic sections of $\oplus^n E \rightarrow \Sigma$ and, using the correspondences of \cite{B-D-W, B-D-GP-W}, relate them to vortex solutions, thereby proving theorem \ref{thm1.1}. Finally in section 4 we look at the space of all vortex internal structures of fixed order $k$, and argue that it is a compact and connected space. We end up by comparing our description with the special cases $k=1,2$ already studied in the literature.

\section{Holomorphic matrix functions on the plane}

Matrix functions $\phi (z)$ defined on the plane $\Sigma = \CC$ were studied at length in \cite{Eto-1}, where they were called moduli matrices. The word moduli appears because what we really want to study are the equivalence classes of $\phi (z)$'s related by the equivalence relation $\phi(z) \sim V(z) \: \phi(z)$, with $V(z)$ any matrix function that is invertible for all $z$. This equivalence can be called a complex gauge transformation or, in the language of \cite{Eto-1}, a V-transformation.

Our approach here is rather different from the one in \cite{Eto-1}. In those articles the classes of functions $\phi(z)$ are characterized by the coefficients of the various polynomials that appear in the matrix; here we characterize them by the position in $\CC$ of the zeros of $\det \phi (z)$ and by the factorization of $\phi (z)$ around each of these zeros. This last method seems to make easier the generalization to compact $\Sigma$. The first step is the following local factorization result.
\begin{prop}
Let $\phi (z)$ be a holomorphic function of one complex variable with values on the square $n\times n $ matrices. Suppose that $\phi (z)$ is defined in a neighbourhood of the origin $z=0$ and that the function $\det \phi (z)$ has an isolated zero of order $k$ at this point. Then there exists a unique internal structure $\is_n (\phi) = (k_0 , V_1 , \ldots , V_l)$ such that $\phi (z)$ can be written around the origin as
\begin{equation}
\phi (z) \ = \ A(z)\: z^{k_0} \: T_{V_l}(z) \cdots T_{V_1} (z) \ ,
\label{2.1}
\end{equation}
where $A(z)$ is a holomorphic matrix function that is invertible around $z=0$. Clearly the order of this $\is_n (\phi)$ is precisely $k$.
\label{prop2.1}
\end{prop}

\begin{cor}
Any internal structure $\is_n$ can be obtained as $\is_n (\phi)$ for an appropriate holomorphic function $\phi (z)$ defined around the origin.
\label{cor2.1}
\end{cor}

\begin{cor}
Two holomorphic functions $\phi_1 (z)$ and $\phi_2 (z)$ determine the same internal structure if and only if $\phi_2 (z) = A(z)\: \phi_1 (z)$ for some invertible matricial function $A(z)$.
\label{cor2.2}
\end{cor}

\begin{prooff}
We start by showing how to obtain the structure $\is_n (\phi)$ from the function $\phi (z)$. The integer $k_0$ is defined as the minimal order of the zeros at the origin of the $n^2$ entries of the matrix $\phi(z)$, or in other words it is the only integer such that $\phi_1 (z) := z^{-k_0} \phi (z)$ is holomorphic and does not vanish at the origin. Observe that $\det \phi_1 (z)$ has a zero of order $k- n k_0$ at $z=0$. If this order is zero, then one takes $l=0$ and no vector spaces appear on $\is_n (\phi)$. If on the other hand $k - nk_0 > 0$, one defines $V_1 = \ker \phi_1 (0)$ as the first proper subspace of $\CC^n$. To continue with the procedure, define the second function
\[
\phi_2 (z)\ := \ \phi_1 (z) \: T_{V_1}^{-1} (z)
\]
By applying this transformation to vectors in $V_1$ and in $V_1^\perp$, it is clear that $\phi_2 (z)$ is well defined and holomorphic around the origin, including at the origin itself. Again, if $\phi_2 (0)$ is invertible one takes $l=1$ and the sequence of vector spaces terminates. If not, one defines $V_2 = \ker \phi_2 (0)$ and the sequence of subspaces continues. The fact that $\phi_2 (0)$ is injective on $V_1^{\perp}$ implies that $V_2 \cap V_1^{\perp} = \{0 \}$. Moreover, it follows from definition \ref{dfn1.1} that $\det \phi_2 (z)$ vanishes at $z=0$ with order $k-nk_0- {\rm dim}\ V_1$. Decomposing once more $\CC^n = V_2 \oplus V_2^{\perp}$ one can go on with the procedure until $k - nk_0 - \sum_{j=1}^l {\rm dim}\ V_j$ vanishes for some $l$. When this happens the sequence of vector subspaces terminates and the linear transformation
\[
\phi_{l+1}(z)\ := \ \phi_l (z)\: T_{V_l}^{-1} (z) \ = \ \phi (z) \: z^{-k_0} \: T_{V_1}^{-1}(z) \cdots T_{V_l}^{-1}(z)
\]
will be invertible at the origin. This shows the existence of $\is_n (\phi)$ and of the decomposition (\ref{2.1}).

For the uniqueness part, suppose that $\phi (z)$  given by (\ref{2.1}) had another decomposition associated to a second set of data $(k'_0 , V'_1 , \ldots , V'_{l'})$. Then the function $z^{-k_0} \phi (z)$ would be locally given by 
\begin{equation} 
A(z)\: T_{V_l} (z) \cdots T_{V_1}(z) \ = \ z^{k'_0 -k_0}\: A'(z)\: T_{V'_{l'}}(z) \cdots T_{V'_1} (z) \ .
\label{2.2}
\end{equation}
The essential point now is that the conditions $V_{j+1} \cap V_j^{\perp} = \{ 0 \}$ imply that for all $s\leq l$ the kernel of
\[
T_{V_l}(0) \cdots T_{V_s} (0) \ = \ \Pi_{V_l}^{\perp} \cdots \Pi_{V_s}^{\perp} 
\]
is $V_s$. This is obvious for $s=l$ and then clear by induction. Thus applying this fact to the left-hand side of (\ref{2.2}), we see that $z^{-k_0} \phi (z)$ is well defined and has kernel $V_1$ at the origin. But then looking at the right-hand side, this can be true only if $k_0 = k'_0$ and $V_1 =V'_1$. Arguing in the same way for the function $z^{k_0} \phi(z) T_{V_1}^{-1}(z)$ we would conclude that also $V_2 =V'_2$, and so forth for the other $V_j$'s. This finishes the proof of the proposition.

As for the first corollary, it is enough to note that given an internal structure $\is_n = (k_0 , V_1 , \ldots , V_l)$ we can just define
\[
\phi(z)\ := \ z^{k_0} \: T_{V_l}(z) \cdots  T_{V_1}(z) \ .
\]
By the uniqueness of the local factorization it is then obvious that for this choice $\is_n (\phi) = \is_n $. The second corollary is a direct consequence of the decomposition (\ref{2.1}).
\end{prooff}
Having understood the local internal structures associated to each zero of $\det \phi$, we will now see how the set of these zeros and internal structures effectively characterizes any matrix function defined globally on $\CC$.
\begin{thm}
Let $\{ z_1 , \ldots , z_r \}$ be any finite set of distinct points in the complex plane and let $\{ \is_n^1 , \ldots , \is_n^r  \}$ by any set of internal structures as defined in \ref{dfn2.2}. Then there exists a holomorphic function $\phi (z)$ with values on the $n \times n$ matrices such that $\det \phi (z)$ has zeros exactly at the points $z_j$ with associated internal structure $\is_n^j$. This function $\phi (z)$ is unique up to left multiplication by globally invertible matrix functions. 
\label{thm2.1}
\end{thm}

\begin{prooff}
The case $r=1$ follows directly from proposition \ref{prop2.1}. Arguing by induction, suppose now that $\det \phi(z)$ has zeros at the points $z_1, \ldots, z_{r-1}$ with associated internal structures $\is_n^1 , \ldots , \is_n^{r-1}$, and that at the point $z=z_r$ the structure of $\phi (z)$ is given by $(k_0, V_1 , \ldots , V_{l-1})$. Given any vector subspace $V_l \subset \CC^n$ such that $V_l \cap V_{l-1}^{\perp} = \{ 0 \}$, we will construct a new function $\tilde{\phi}(z)$ with the same structure as $\phi (z)$ at the points $z_1 , \ldots , z_{r-1}$ and, at the point $z=z_r$, internal structure $(k_0, V_1 , \ldots , V_{l-1}, V_l)$. This will suffice to prove the existence part of the theorem.

To start with, by proposition \ref{prop2.1} we have that $\phi (z)$ can be written around $z_r$, and hence globally, as
\[
\phi (z) \ = \ A(z)\: (z-z_r)^{k_0} \: T_{V_{l-1}} (z-z_r) \cdots T_{V_1} (z-z_r)
\]
for some $A(z)$ invertible around $z_r$. Now define the new function
\[
\tilde{\phi}(z)\ := \ T_{V_l} (z-z_r) \: A(z_r)^{-1} \: \phi (z) \ .
\]
Since $\tilde{\phi}(z)$ is related to $\phi (z)$ by left multiplication by a matrix invertible around $z_1 , \ldots , z_{r-1}$, the internal structures of $\tilde{\phi}(z)$ and $\phi (z)$ at these points are the same. Furthermore, observe that the matrix $B(z) = T_{V_l} (z-z_r) \: A(z_r)^{-1} \: A(z)$ has vanishing determinant at $z=z_r$ with order ${\rm dim}\ V_l$ and that the kernel of $B(z_r)$ is exactly $V_l$. It then follows from proposition \ref{prop2.1} that $B(z)$ can be written as $C(z)\: T_{V_l} (z-z_r)$ for some $C(z)$ invertible around $z_r$. In particular one can also rewrite
\[
\tilde{\phi}(z)\  = \ C(z)\: (z-z_r)^{k_0} \: T_{V_l}(z-z_r)\: T_{V_{l-1}} (z-z_r) \cdots T_{V_1} (z-z_r)\ .
\]
Appealing again to proposition \ref{prop2.1} one concludes that the internal structure of $\tilde{\phi}(z)$ at $z=z_r$ is $(k_0, V_1 , \ldots , V_l)$, as we wanted.

For the uniqueness, suppose that both $\phi (z)$ and $\phi' (z)$ satisfy the conditions of the theorem. Then both $\phi (z)$ and $\phi' (z)$ can be factorized as in (\ref{2.1}) around any of the $z_j$'s with the same internal structure $\is_n^j$; only the matrices $A(z)$ in (\ref{2.1}) will possibly differ. But then it is clear that the matrix $\phi' (z) \phi^{-1}(z)$ is well defined and invertible around $z_j$. Since this happens for all of the $z_j$'s we conclude that $\phi' (z) \phi^{-1}(z)$ is globally well defined and invertible. Finally the tautology $\phi' (z)= [\phi' (z) \phi^{-1}(z)]\: \phi (z)$ concludes the proof.
\end{prooff}

\vspace{-0.2cm}

\section{Non-abelian vortices on compact Riemann surfaces}

In the first proposition we extend the results of section 2 to compact $\Sigma$, i.e. to holomorphic sections of holomorphic bundles over $\Sigma$. After that we use the Hitchin-Kobayashi-type correspondences of \cite{B-D-W, B-D-GP-W} to relate these holomorphic sections to the actual vortex solutions over $\Sigma$, thereby proving theorem \ref{thm1.1}. Finally at the end of the section we state a result that gives a more practical interpretation of the constant $\tau$ that appears in the vortex equations. The proof is omited, because it is almost identical to the $n=1$ case proved in \cite{Brad}.  

\begin{prop}
Let $\Sigma$ be a compact Riemann surface, let $\{ z_1 , \ldots , z_r \}$ be any set of distinct points in the surface and let $\{ \is_n^1 , \ldots , \is_n^r  \}$ be any set of internal structures. Then there exists a holomorphic vector bundle $E\rightarrow \Sigma$ of rank $n$ and a section $\phi$ of $\oplus^n E$ such that $\det \phi$ has zeros exactly at the points $z_j$ with respective internal structure $\is_n^j$. This pair $(E,\phi)$ is unique up to isomorphisms of holomorphic bundles. Moreover, the degree of the bundle $E$ is equal to the sum of the orders of the $\is_n^j$'s.
\label{prop3.1}
\end{prop}

\begin{prooff}
It is always possible to take a connected open set ${\mathcal V}_0$ of $\Sigma$ that contains all the $z_j$'s and is simultaneously the domain of a complex chart of the surface. Now, using theorem \ref{thm2.1}, construct a holomorphic matrix function $\tilde{\phi}(z)$ on the open set ${\mathcal V}_0$ with zeros at the points $z_j$ and respective internal structure $\is_n^j$. Considering the complementary set ${\mathcal V}_1 = \Sigma \setminus \{ z_1 , \ldots , z_r \}$, we have that the restriction of $\tilde{\phi}(z)$ to the intersection ${\mathcal V}_0 \cap {\mathcal V}_1$ --- which we call $\psi$ --- is a $n\times n$ invertible matrix on this set, and so can be taken as the transition function for some holomorphic vector bundle $E \rightarrow \Sigma$ of rank $n$ that is trivial over ${\mathcal V}_0$ and ${\mathcal V}_1$. Finally, taking simultaneously the constant function $1_{n\times n}$ on ${\mathcal V}_1$ and the function $\tilde{\phi}$ on ${\mathcal V}_0$, we have that over the intersection of the sets they satisfy the compatibility condition $\tilde{\phi}(z) = \psi (z) \: 1_{n\times n}$, and so these two functions define a global section $\phi$ of the direct sum $\oplus^n E$. It is clear that this $\phi$ has the required properties.

To prove uniqueness, suppose that $(E' , \phi')$ was another pair with the required properties. Then taking local holomorphic trivializations of $E$ and $E'$ and representing the sections $\phi$ and $\phi'$ by local holomorphic matrix functions, corollary \ref{cor2.2} implies that the matrix $\phi' \phi^{-1}$ is holomorphic and invertible throughout the whole domain of trivialization. Moreover, if we pick different trivializations of $E$ and $E'$ related to the initial ones by transition functions $s$ and $s'$, then the matrix $\phi' \phi^{-1}$ obviously transforms as $s' \phi' \phi^{-1}s^{-1}$, which implies that the matrices $\phi' \phi^{-1}$ actually define a global, holomorphic and invertible section of ${\rm Hom}(E, E')\rightarrow \Sigma$, or in other words an isomorphism $\Lambda: E\rightarrow E'$. Since clearly $\Lambda (\phi)=\phi'$, the uniqueness is proved.

Finally, to justify the last statement, just observe that the section $\det \phi$ of the determinant bundle $\det E \rightarrow \Sigma$ vanishes exactly at the points $z_j$ with multiplicity equal to the order of the respective $\is_n^j$. Well-known properties of line-bundles then imply that $\det E$, and hence $E$, have degree equal to the sum of the orders of the $\is_n^j$'s.
\end{prooff}

 After studying the holomorphic part of the problem, i.e. after constructing the holomorphic vector bundles and the holomorphic sections with the required zeros and internal structures, we are now in position to relate them to the actual solutions of the vortex equations. 
\vspace{0.3cm}

\begin{prooff1.1}
In short, the key point here comes from the results of \cite{B-D-W, B-D-GP-W}, which guarantee that under the volume assumption $({\rm Vol}\; \Sigma) > 2\pi d / (e^2\, \tau)$ each pair $(E,\phi)$ constructed in proposition \ref{prop3.1} is in fact a $\tau$-stable $n$-pair, and hence there is a complex gauge transformation that takes it to a solution of the vortex equations.

To understand more clearly what this means, suppose that we are given a $C^\infty$ vector bundle $E$ and a finite set of pairs $(z_j , \is_n^j)$ satisfying the conditions of theorem \ref{thm1.1}. Then by the proposition above there is a complex structure on $E$ and a holomorphic section $\phi$ of $\oplus^n E$ such that $\det \phi$ vanishes at the $z_j$'s with internal structure $\is_n^j$. Using the fixed hermitian metric $h$ on $E$, this complex structure on its turn defines a natural connection $A$ on $E$ --- the so-called Chern connection --- such that $\bar{\partial}^A \phi = 0$. We thus have a solution $(A, \phi)$ of the first vortex equation. Observe now that while this first equation is invariant under complex gauge transformations, i.e. gauge transformations with the complexification $U(n)^\CC = SL(n, \CC)$ as the gauge group, the second vortex equation is certainly not invariant. This entitles us to ask whether a complex gauge transformation can take our pair $(A,\phi)$, which a priori only satisfies the first equation, to a solution of also the second equation, and hence to a full vortex solution. The answer is that this is possible whenever the initial holomorphic $(E, \phi)$ is a $\tau$-stable $n$-pair, and that in this case the required complex gauge transformation is unique up to real gauge transformations. We thus see how the results of \cite{B-D-W, B-D-GP-W} give a vortex solution for each $(E, \phi)$ constructed in proposition \ref{prop3.1}.

To end the proof of theorem \ref{thm1.1} there are a few more points that should be checked. The first is to note that complex gauge transformations on $(A,\phi)$ do not change the holomorphic structure on $E$ induced by $A$ (up to equivalence) and that, furthermore, they transform $\phi$ through holomorphic isomorphisms of the bundles. This means that in the local factorization (\ref{1.3}) the only thing that changes under complex gauge transformations is the matrix $A(z)$, and hence the vortex solutions that were obtained in the last paragraph by means of complex transformations have exactly the same internal structures $\is_n^j$ at the zeros $z_j$ as the solutions $\phi$ constructed in proposition \ref{prop3.1}.

The second thing to check is the uniqueness. If we were given two vortex solutions with the same zeros of $\det \phi$ and internal structures, then, as described in the introduction, we would get two complex structures on $E$ and respective holomorphic sections with the same zeros and $\is_n$'s. From the uniqueness in proposition \ref{prop3.1}, however, it follows that these two holomorphic pairs $(E, \phi)$ would in fact be isomorphic. Finally from the uniqueness in the results of \cite{B-D-W, B-D-GP-W} for the required complex gauge tranformations, we get that the initial vortex solutions must have been related by a real gauge transformation.

The last point to check is the final assertion of theorem \ref{thm1.1}. Here the justification comes from the fact that, as described in the introduction, to any given vortex solution on $\Sigma$ with $\det \phi \neq 0$ we can associate a finite set of pairs $(z_j ,\is_n^j)$'s. Using these pairs to construct our own vortex solution according to the prescriptions of this section, the uniqueness part in theorem \ref{thm1.1}, which is already proved, assures us that the given vortex solution is in fact gauge equivalent to the constructed one.
\end{prooff1.1}

Another interesting  property of the non-abelian vortex solutions, proved as in \cite{Brad}, is the following.
\begin{prop}
Let $(A, \phi)$ be a solution of the vortex equations (\ref{1.2}) on a $C^\infty$ hermitian bundle $(E, h)$ of rank $n$. Then the norm of each of the $n$ components $\phi_j$ of the section the section $\phi : \Sigma \rightarrow \oplus^n E$ satisfies the majoration $|\phi_j (z)|^2_h \leq \tau$ for all $z$ in $\Sigma$.   
\end{prop}

\section{On the space of vortex internal structures}

\subsection{Connectedness and compactness}

In the introduction to this paper we defined an internal structure $\is_n$ as a set consisting of an integer plus a sequence of vector subspaces of $\CC^n$ satisfying a non-intersection condition. The advantages of this characterization are that it is simple enough, general for all $n$ and, through the local factorization (\ref{1.3}), directly related to the actual behaviour of the vortex solutions. A significant disadvantage is that it does not reveal transparently the main geometric and topological properties of the space of all internal structures, i.e. of the internal configuration space of the vortex. For example a priori one could think that the non-intersection condition would imply that these internal spaces are non-compact, a fact that is not true. In this final section of the paper we will spend some time looking at these matters, and will conclude that in a natural topology the space $\ais_{n,k}$ of all internal structures of fixed order $k$ is in fact compact and connected.

  To begin with observe that one can divide the space $\ais_{n,k}$ into disjoint strata according to the dimensions of the vector spaces at each point $\is_n = (k_0 , V_1 , \ldots , V_l)$. Calling $k_j$ the dimension of $V_j$, the fixed order condition means that $nk_0 + k_1 + \cdots + k_l = k$, and then each vector of integers $(k_0 , k_1 , \ldots , k_l)$ satisfying this equality labels exactly one of these strata. Now inside each stratum, by definition, the dimensions of $V_j$ are fixed, so the only degrees of freedom are the spacial orientations of these subspaces inside $\CC^n$, and these are parametrized by the grassmannian ${\rm Gr}(k_j , n)$. Note however that due to the non-intersection conditions not all spacial orientations are allowed, and the degrees of freedom are parametrized only by a dense open subset of the grassmannian. This implies that in general the different strata are not necessarily compact, although as we will see the full space $\ais_{n,k}$ is.

Now, the complex dimension of a stratum labelled by $(k_0 , \ldots , k_l)$ is of course $\sum_{j=1}^l  k_j (n-k_j)$, where each term in the sum is the dimension of a grassmannian. One can then show that, under the constraint imposed by the fixed order, this sum is strictly maximized with the choice $k_0 = 0$, $l=k$ and $k_j = 1$ for $j\geq 1$, where it has value $k(n-1)$. This means that the highest dimensional stratum in $\ais_{n,k}$, i.e. the generic part of the internal space, can be discribed as a choice of $k$ lines in $\CC^n$, or in other words as a dense open subset of the $k$-fold cartesian product of $\CC{\mathbb P}^{n-1}$. Moreover, we will see how the complement in $\times^k\, \CC{\mathbb P}^{n-1}$ of this dense open subset in fact describes all the remaining strata of $\ais_{n,k}$, although different points in the complement can represent the same non-generic point of $\ais_{n,k}$. More precisely we have the following result.
\begin{prop}
There is a natural surjective map $\mais : \times^k\, \CC{\mathbb P}^{n-1} \rightarrow \ais_{n,k}$ that is one-to-one on the dense open subset of $\times^k\, \CC{\mathbb P}^{n-1}$ defined by the multiplets $(L_1, \ldots , L_k)$ that satisfy the non-intersection condition $L_{j+1} \cap L_j^\perp = \{ 0\}$. Since this map is surjective, the set of internal structures $\ais_{n,k}$ equipped with the quotient topology is a compact and connected space. 
\label{prop4.1}
\end{prop}         
\begin{prooff}
The argument for the proof is quite simple and goes as follows. Let $L = (L_1 , \ldots, L_k )$ be a point in $\times^k\, \CC{\mathbb P}^{n-1}$. Irrespective of whether $L$ satisfies or not the non-intersection condition, it makes sense to consider the the linear transformations
\begin{equation*}
\hat{T}_L (z) \ := \ T_{L_k} (z)  \cdots T_{L_1}(z)
\end{equation*} 
for all complex $z$. It is clear that $\hat{T}_L (z)$ is holomorphic and that $\det \hat{T}_L (z)$ has a single zero at $z=0$ of vanishing order $k$. Thus applying theorem \ref{thm2.1} there is a unique internal structure $\is_n = (k_0 , V_1 , \ldots,  V_l)$ of order $k$ such that 
\begin{equation*}
\hat{T}_L (z) \ = \ z^{k_0}\: T_{V_l}(z) \cdots T_{V_1}(z) \ ,
\end{equation*}
and this defines the map $\mais$.

Now call ${\mathcal U}_0$ the set of $L$'s in $\times^k\, \CC{\mathbb P}^{n-1}$ that satisfy the non-intersection condition, i.e. the set of $L$'s that are proper internal structures. It is clear that $\mais$ restricted to ${\mathcal U}_0$ is injective with image the generic stratum of $\ais_{n,k}$; this is a tautology, for in fact $\mais$ restricted to ${\mathcal U}_0$ is the identity. Outside ${\mathcal U}_0$, on the other hand, the story is different, for if $L_{j+1} \cap L_j^\perp$ is non-zero then $L_{j+1} \subseteq  L_j^\perp$, and it is easy to check that in this case
\begin{equation}
T_{L_{j+1}} (z) \ T_{L_j} (z) \ = \ T_{L_{j+1} \oplus L_j} (z) \ .
\label{4.1}
\end{equation}
This equality is a special instance of the more general lemma \ref{lem4.1}. It means that if $L_{j+1}$ and $L_j^\perp$ have non-zero intersection the map $\mais$ only depends on the direct sum $L_{j+1} \oplus L_j$, and so is clearly not injective.

The identities (\ref{4.1}) and (\ref{4.2}) can also be applied recursively to show that for any vector subspace $V$ of $\CC^n$ there exist lines $L_1 ,  \ldots , L_{\dim V}$ such that 
\begin{equation*}
T_V (z) \ = \ T_{L_{\dim V}}(z) \cdots T_{L_1} (z) \ . 
\end{equation*}
In fact any set of orthogonal lines such that $V = L_1 \oplus \cdots \oplus L_{\dim V}$ will do the job. This shows that $\mais$ is a surjective map. 
\end{prooff}

\begin{lem}
Let $V_1$ and $V_2$ be any two subspaces of $\CC^n$. Then calling $W$ the intersection $V_2 \cap V_1^\perp$, the linear transformations of definition \ref{dfn1.1} satisfy the algebraic identity
\begin{equation}
T_{V_2}(z)\ T_{V_1}(z) \ =\ T_{V_2 \, \cap \, W^\perp }(z) \  T_{W \oplus V_1} (z) \ .
\label{4.2}
\end{equation}
Observe moreover that the two subspaces on the right-hand side satisfy the usual non-intersection condition, i.e. $V_2 \cap W^\perp$ has zero intersection with $(V_1 \oplus W)^\perp$.
\label{lem4.1}
\end{lem}
\begin{prooff}
It is clear from definition \ref{dfn1.1} that the identity above is equivalent to the three separate identities
\begin{align}
&{\rm (i)} \quad \Pi_{V_2} \ \Pi_{V_1}\ = \ \Pi_{W^\perp \cap V_2} \ \Pi_{W\oplus V_1} \ ;   \nonumber  \\
&{\rm (ii)}  \quad \Pi^\perp_{V_2} \ \Pi^\perp_{V_1}\ = \ \Pi^\perp_{W^\perp \cap V_2} \ \Pi^\perp_{W\oplus V_1} \ ;  \nonumber \\
&{\rm (iii)} \quad  \Pi^\perp_{V_2} \ \Pi_{V_1} \ + \ \Pi_{V_2} \ \Pi^\perp_{V_1}\ \ = \ \Pi^\perp_{W^\perp \cap V_2} \ \Pi_{W\oplus V_1}\ +\  \Pi_{W^\perp \cap V_2} \ \Pi^\perp_{W\oplus V_1}\ .   \nonumber
\end{align}
Since the proofs of these equalities are rather cumbersome and similar to each other, here we will restrict ourselves to prove (i).

The first step is to note that the general identity of subspaces $(A+B)^\perp = A^\perp \cap B^\perp$ implies that each of the decompositions
\begin{align}
\CC^n \ &= \  V_1 \oplus W \oplus (W^\perp \cap V_1^\perp)  \label{4.3} \\
        &= \ V_2^\perp \oplus W \oplus (W^\perp \cap V_2)\  \label{4.4}
\end{align}
is orthogonal. Having this in mind, suppose now that $v$ is a vector in $W$. Then we have that
\[
\Pi_{W^\perp \cap V_2} \ \Pi_{W\oplus V_1}\ (v) \ = \ \Pi_{W^\perp \cap V_2} \ (v) \ = \ 0 \ =\ \Pi_{V_1} \ (v) \ ,
\]
and so both sides of (i) annihilate $v$. If on the other hand $v$ belongs to the subspace $W^\perp \cap V_1^\perp = (W\oplus V_1)^\perp$, we have that 
\[
\Pi_{W\oplus V_1} \ (v)\ = \ 0\ =\ \Pi_{V_1} \ (v)\ ,
\]
and so also in this case both sides of (i) annihilate $v$. Finally suppose that $v$ sits in $V_1$. It is clear that
\[
\Pi_{W\oplus V_1}\ (v)\ =\ v\ =\ \Pi_{V_1}\ (v)\ ,
\]
and hence we only have to check that
\[
\Pi_{V_2}\ (v) \ = \Pi_{W^\perp \cap V_2} \ (v) \ .
\]
But $V_1$ being orthogonal to $W$, we can use the decomposition (\ref{4.4}) to write $v = v_2 + w_2$ with $v_2 \in V_2^\perp$ and $w_2 \in W^\perp \cap V_2$. Then obviously $\Pi_{W^\perp \cap V_2} (v) = w_2$. Now since $\Pi_{V_2} (v_2) = 0$ and $w_2\in  V_2$, we finally have that $\Pi_{V_2} (v) = \Pi_{V_2} (w_2) = w_2$, which concludes the proof of (i). 
\end{prooff}

\subsection{The case $k=2$: comparison with the literature}

The space of vortex internal structures has been described previously in the literature in the cases $k=1$ and $k=2$. The case $k=1$ is simple enough: it follows directly from definition \ref{1.2} that $\ais_{n,1} \simeq \CC {\mathbb P}^{n-1}$, as was known. The case $k=2$ has been studied in \cite{A-S-Y, Hash-T} for $n=2$ and in \cite{Eto-4} for general $n$. Here we will compare our results with those of \cite{Eto-4}, finding in the end that they are consistent.

We start with our results. According to the previous subsection the highest dimensional stratum of $\ais_{n,2}$ has dimension $2(n-1)$ and isomorphic to
\[
{\mathcal U}_0 \ = \ (\CC {\mathbb P}^{n-1} \times \CC {\mathbb P}^{n-1}) \setminus S \ ,
\] 
where $S$ is the submanifold defined by $L_2 \cap L_1^\perp \neq \{ 0 \}$, or in other words by $L_2 \perp L_1$. This is the domain ${\mathcal U}_0$ where the map $\mais$ of proposition \ref{prop4.1} is injective. Outside ${\mathcal U}_0$, i.e. on $S$, the line $L_2$ is perpendicular to $L_1$, and so it follows from (\ref{4.1}) that in this case $\mais (L_1, L_2) = L_1 \oplus L_2 \subseteq \CC^n$, i.e. the internal structure associated to $(L_1 , L_2) \in S$ is the 2-dimensional vector space $L_1 \oplus L_2$. Since these vector spaces are parametrized by the grassmannian ${\rm Gr}(2, n)$ we recognize that, in informal terms, $\ais_{n,2}$ is isomorphic to $\CC {\mathbb P}^{n-1} \times \CC {\mathbb P}^{n-1}$ with the submanifold $S$ collapsed into the grassmannian ${\rm Gr}(2,n)$.   
 
(Incidentally, identifying the orthogonal space $L^\perp$ with the tangent space to $\CC {\mathbb P}^{n-1}$ at the point $L$, it is manifest that the submanifold $S$ is itself isomorphic to the projectivization of the tangent bundle $T \CC {\mathbb P}^{n-1}$, which is a bundle over $\CC {\mathbb P}^{n-1}$ with fibre $\CC {\mathbb P}^{n-2}$.)

In the paper \cite{Eto-4}, on the other hand, the space of vortex internal structures for $k=2$ was described in the following terms. Let ${\mathcal M}$ be the space of $2\times (n+1)$ matrices of rank 2. We can write every element $M \in {\mathcal M}$ in the form
\[
M\ = \ \begin{bmatrix}
        \psi_2^T  &  v_2 \\
        \psi_1^T  &  v_1    \end{bmatrix} \ = \ [\, \psi \quad v\, ] \ , 
\]
with $\psi_1 , \psi_2 \in \CC^n$ and $v \in \CC^2$. Consider now the left action of the group $\CC^\ast \times SL(2 , \CC)$ on ${\mathcal M}$ defined by 
\begin{equation}
(\lambda , A)\cdot [\,\psi \quad v\,]\ = \ [\,\lambda A \psi \quad Av\,] \ .
\label{4.5}
\end{equation}
Then according to \cite{Eto-4} there is an isomorphism
\begin{equation}
\ais_{n,2} \ \simeq \ {\mathcal M}\ / \ \CC^\ast \times SL(2, \CC) \ . 
\label{4.6}
\end{equation}
To compare this result with our previous description we will study in detail the equivalence classes in this quotient.

Firstly, under the action of $\CC^\ast \times SL(2 , \CC)$ one can distinguish two types of orbits in ${\mathcal M}$: those with $v=0$ and those with $v\neq 0$. The points in the orbits with $v=0$ have linearly independent $\psi_1$ and $\psi_2$, for $M$ must have rank 2. When $\CC^\ast \times SL(2 , \CC)$ acts on these points the only invariant is the subspace $V = {\rm span}_\CC \{\psi_1 ,\psi_2  \}$, and these subspaces are exactly parametrized by the grassmannian ${\rm Gr}(2, n)$. Thus decomposing ${\mathcal M} = {\mathcal M}_{v=0} \cup {\mathcal M}_{v\neq 0}$ we see that 
\begin{equation*}
{\mathcal M}_{v=0} \ / \ \CC^\ast \times SL(2 , \CC) \ = \ {\rm Gr}(2,n) \ ,
\end{equation*}
which is part of the internal space described in our results. Now we want to prove that there is also an isomorphism
\begin{equation}
\Lambda\ :\ {\mathcal M}_{v\neq 0} \ / \ \CC^\ast \times SL(2 , \CC)\ \longrightarrow \ (\CC {\mathbb P}^{n-1} \times \CC {\mathbb P}^{n-1})\setminus S \ ,
\label{4.7}
\end{equation}
and so conclude that there exists a bijection between the full internal spaces $\ais_{n,2}$ as described here and in \cite{Eto-4}.

So consider the equivalence classes of the points $[\,\psi \quad v\,]$ with $v\neq 0$. Since $SL(2, \CC)$ acts transitively on $\CC^2 \!\setminus \! \{ 0\}$, any such equivalence class has a representative of the form
\begin{equation}
\begin{bmatrix}
        \psi_2^T  &  1 \\
        \psi_1^T  &  0    \end{bmatrix} \ .
\label{4.8}
\end{equation}
Observe that here $\psi_1 \neq 0$, since the rank of the matrix must be 2. The representative (\ref{4.8}), however, is not unique; on the one hand because it can be acted upon by the $\CC^\ast$ subgroup, and on the other hand because the vector $[1\quad 0]^T$ has a one dimensional stabilizer under the $SL(2, \CC)$-action. All in all, the representative is unique up to transformations of the form
\begin{equation}
\begin{bmatrix}
        \psi_2^T  &  1 \\
        \psi_1^T  &  0    \end{bmatrix}\ \longmapsto \ \lambda \ 
\begin{bmatrix}
        1  &  a \\
        0  &  1    \end{bmatrix}  \cdot   
\begin{bmatrix}
        \psi_2^T  &  1 \\
        \psi_1^T  &  0    \end{bmatrix} \ = \ 
\begin{bmatrix}
        \lambda (\psi_2 + a\psi_1)^T  &  1 \\
        \lambda\, \psi_1^T  &  0    \end{bmatrix} \  ,
\label{4.9}
\end{equation} 
where $\lambda \in \CC^\ast$ and $a \in \CC$. Now notice that the complex lines in $\CC^n$
\begin{align}
L_1 \ &:= \ {\rm span}_\CC \{ \psi_1\}    \label{4.10} \\
L_2 \ &:= \ {\rm span}_\CC \{ \psi_1 + \Pi_{L_1}^\perp (\psi_2) \} \nonumber
\end{align}
are well defined and are invariants of the transformations (\ref{4.9}). Moreover, it is not difficult to recognize that these lines actually distinguish the orbits of these transformations, i.e. that two matrices of the form (\ref{4.8}) that define different lines cannot be related by a transformation (\ref{4.9}), and hence lie on different $\CC^\ast \times SL(2, \CC)$ orbits. To sum up, through the representative (\ref{4.8}) and the definition (\ref{4.10}) we have constructed an injective map that takes any orbit in ${\mathcal M}_{v\neq 0} \ / \ \CC^\ast \times SL(2 , \CC)$ to a pair of complex lines $(L_1 ,L_2)$. These lines are clearly non-orthogonal, and it is manifest that any pair of non-orthogonal lines can be obtained through (\ref{4.10}) for an appropriate choice of $\psi$. This finally shows that the injective map that we constructed has image $(\CC {\mathbb P}^{n-1} \times \CC {\mathbb P}^{n-1}) \! \setminus\! S$, and hence is the isomorphism (\ref{4.7}) that we were seeking.

As a curiosity, the inverse of the isomorphism $\Lambda$ can be written down very simply as
\[
(L_1 , L_2) \ \longmapsto \ {\rm equivalence\ class\ of}\
\begin{bmatrix}
        \Pi_{L_1}^\perp (w_2)  &  1 \\
        \Pi_{L_1} (w_2)  &  0    
\end{bmatrix} \ ,
\]
where $w_2$ is any non-zero vector in $L_2$. Furthermore, picking any non-zero vector $w_1$ in $L_1$ and decomposing $w_2 = \alpha w_1 + \Pi_{L_1}^\perp (w_2)$, with $\alpha$ the complex scalar $\bar{w_1}^T w_2  / |w_1|^2$, even the full isomorphism
\[
\tilde{\Lambda}\ : \ \ais_{n,2} \ = \ (\CC {\mathbb P}^{n-1} \times \CC {\mathbb P}^{n-1}))\; /\; \mais \ \longrightarrow \ {\mathcal M} \; / \; \CC^\ast \times SL(2 , \CC)
\]
can be written down in the single expression
\[
\tilde{\Lambda}\ : \ {\rm equiv.\ class\ of}\ (L_1 ,L_2) \ \ \longmapsto \ \ {\rm equiv.\ class\ of}\ 
\begin{bmatrix}
 \Pi_{L_1}^\perp (w_2)  &  \sqrt{\alpha} \\
   w_1   &  0 
\end{bmatrix}
\]
for all $L_1 ,L_2$ in $\CC {\mathbb P}^{n-1}$. To check that $\tilde{\Lambda}$ is really well defined and does not depend on the sign choice in $\sqrt{\alpha}$, one must of course make use of the $\CC^\ast \times SL(2, \CC)$-equivalences.

We also note that in general the internal spaces $\ais_{n,k}$, although well-defined compact topological spaces, are not necessarily smooth manifolds outside of the highest dimensional stratum. This has been illustrated at length in \cite{Eto-4} in the case $k=2$. In the discussion above a reflection of this singularity is the appearance of $\sqrt{\alpha}$ in the isomorphism $\tilde{\Lambda}$.

\vskip 25pt
\noindent
{\bf Acknowledgements.}
It is a pleasure to thank Minoru Eto, Muneto Nitta, Keisuke Ohashi and Norisuke Sakai for kindly hosting me during a visit to TITech, Tokyo, three years ago. I am grateful to them and to David Tong for explaining me their work on non-abelian vortices. I am partially supported by the Netherlands Organisation for Scientific Research (NWO) through the Veni grant 639.031.616.

\end{document}